\documentclass[11pt]{article}
\usepackage[margin=1in]{geometry}
\usepackage{amsmath,amssymb,booktabs,graphicx,caption,array,longtable,threeparttable}
\usepackage[backend=biber,style=authoryear,natbib=true,maxcitenames=2,maxbibnames=20]{biblatex}
\usepackage{microtype}
\usepackage[hidelinks]{hyperref}
\usepackage{xspace}
\usepackage{float}

\newcommand{\FIFO}{FIFO\xspace}
\newcommand{\RecentN}{Recent-$n$\xspace}
\newcommand{\HybridN}{Hybrid-$n$\xspace}
\newcommand{\Rare}{Rare replay\xspace}
\newcommand{\UB}{unbounded current-state reference\xspace}

\title{Aggregate Calibration Can Mask Stress-State Undercoverage:\\
Bounded-Memory Forecast Intervals under Recurrent Financial Stress}
\author{Ruihang Han$^{a}$ and Li-Hsiang Lin$^{a}$\\[4pt]
\small $^{a}$Department of Mathematics and Statistics, Georgia State University, Atlanta, GA 30303}
\date{}

\begin{document}
\maketitle

\begin{abstract}
Sequential forecast intervals are often evaluated by aggregate coverage even when calibration memory is bounded. This can be misleading when economically important states recur after long absences: a recent-only buffer may satisfy its overall coverage target while no longer containing errors from earlier episodes of the returning state. We study this problem in next-day financial range forecasting, holding the forecasting model, conformity score, adaptive update, and memory budget constant while varying the historical residual information available at each query. Recent-only first-in, first-out (FIFO) calibration is near nominal in aggregate but undercovers substantially during high-volatility states in both an exchange-traded fund development panel and a fixed stock panel. An unbounded state-history reference yields a lower observed stress-period interval score; Rare replay yields lower high-VIX interval scores than the matched Recent-n and Hybrid-n controls in both panels, and the state-structured versus recent-only contrast is largest near stress re-entry. Bounded calibration memory is therefore part of the forecasting information design, and aggregate calibration alone can conceal state-specific reliability failures.
\end{abstract}

\noindent\textbf{Keywords:} Conformal prediction; Forecast intervals; Financial forecasting; Time series; Uncertainty quantification

\section{Introduction}
\label{sec:intro}

Financial stress is episodic and recurrent. A high-volatility state may return after a long calm period, so forecast errors from an earlier stress episode can remain relevant after many more recent observations have accumulated. For interval forecasting under recurrent market conditions, relevance is therefore not necessarily monotone in recency. A system can look well calibrated when performance is averaged over a long evaluation period while providing substantially lower coverage in the less frequent states in which uncertainty is most consequential.

This creates a specific problem when sequential calibration operates under bounded memory. A first-in, first-out (FIFO) buffer removes residuals according to age, without regard to whether the state in which they were observed may recur. When stress returns, the buffer may no longer contain evidence from the previous episode. If calm dates dominate the evaluation sample, aggregate coverage can remain close to nominal while reliability deteriorates in the returning stress state. We call this retention-induced availability failure \emph{FIFO forgetting}: content-blind eviction changes the information set available for future calibration and can conceal state-specific failure behind satisfactory aggregate performance.

The forecasting literature already treats the amount and composition of historical information as a substantive design choice. Rolling and expanding estimation windows can produce different out-of-sample forecasts, and recent volatility-forecasting studies show that performance can depend nonmonotonically on the amount of retained history \citep{feng2024rolling,feng2025window}. Prediction-interval work likewise emphasizes calibration and sharpness under realistic time-series dependence and model misspecification \citep{shang2023intraday,yang2024empirical,chu2026robust}. Recent research also shows why unconditional interval coverage alone is insufficient for assessing conditional calibration \citep{allen2026interval}. These results motivate a parallel question for uncertainty calibration: when the calibration history itself is bounded, which past forecast errors remain available when a previously observed state returns?

Most online conformal methods intervene after a calibration set is available. Classical conformal prediction provides marginal coverage under exchangeability or related symmetry conditions \citep{vovk2005algorithmic,lei2018distribution}. Under temporal dependence or distribution shift, sequential methods adapt the miscoverage level, residual quantile, or feedback controller \citep{xu2021dynamic,gibbs2021adaptive,gibbs2024online,zaffran2022adaptive,xu2023sequential,bhatnagar2023improved,angelopoulos2023pid}. Rolling calibration sets and alternative updating schedules have also been studied under changing environments \citep{vasiloudis2019quantifying,werner2021updating}. This literature does not directly isolate recurring-state evidence loss under a shared hard memory budget with matched information-set controls.

Weighted, localized, and retrieval-based conformal methods address a complementary problem: how the observations that remain available should be used. Such procedures reweight or select historical calibration scores using covariate, local, learned-state, or temporal similarity \citep{tibshirani2019covariate,barber2023beyond,guan2023localized,hore2025conformal,auer2023hopfield,lee2025kernel,neglia2026rescp}. State-Matched Sequence Conformal Prediction (SMSCP) constructs a pre-target state and selects a target-specific reference set of past calibration scores by state similarity, allowing temporally distant observations to remain relevant when comparable states recur \citep{han2026smscp}; Mondrian conformal prediction instead calibrates within predefined categories \citep{bostrom2021mondrian}. These approaches make historical relevance more flexible than pure recency. We focus instead on how eviction under a binding memory budget determines which evidence remains available for later adaptation, localization, or weighting.

The distinction between availability and use also appears in replay and memory systems. Generic reservoir sampling provides a bounded alternative to FIFO storage \citep{vitter1985reservoir}, and replay mechanisms are widely used in continual learning to preserve information that would otherwise be overwritten or forgotten \citep{rolnick2019experience,aljundi2019online}. Recent conformal and uncertainty-calibration work has begun to use retained or replayed buffers across changing tasks and models \citep{gao2025model,shihab2026continual,tumu2026adaptnc}. Here replay varies the calibration information set under a shared memory budget while the forecasting model and conformity score remain fixed.

The financial setting makes this distinction observable. Daily high--low ranges are established measures of price variability \citep{parkinson1980range,chou2005carr}, and the heterogeneous autoregressive model provides a standard long-memory volatility backbone through daily, weekly, and monthly components \citep{corsi2009har}. State-Local Prediction (StaLoP) uses target-local predictive states to borrow from historically distant but locally compatible panels, showing in dynamic panel prediction that temporal proximity need not coincide with predictive relevance \citep{han2026similaritybasedpredictiondigitaltwins}. Recent financial prediction work has considered quantile estimation, similarity-based conformal prediction, and multivariate predictive distributions \citep{li2024quantile,li2025similarity,cheng2025predictive}. A particularly close comparison uses time decay and regime similarity to reweight conformal calibration for nonstationary value-at-risk \citep{schmitt2026tail}. That line changes how available residuals affect the cutoff. We investigate how bounded memory affects the availability of historical residuals from recurrent states, independently of how the retained residuals are subsequently weighted.

We study this question in a controlled next-day financial range-forecasting system. The HAR point forecast, conformity score, adaptive coverage update, memory budget, validation period, and evaluation calendar are held fixed across the main comparisons. Memory construction determines which residuals are stored and partitioned, retrieval determines which stored records become active, and the calibration rule determines how active records affect the next cutoff. Because weighting methods can only use retained observations, their performance may depend on which historical residuals remain available under the memory constraint. The main empirical target is FIFO's stress-state reliability gap, measured as high-state minus non-high-state coverage on the common evaluation calendar; paired high-state interval-score contrasts are then used to compare the coverage--width trade-off across calibration information sets.

The empirical design separates historical availability from simpler explanations. An unbounded current-state reference assesses whether residuals discarded by FIFO remain useful when stress returns. A bounded state-structured replay design then preserves a recent queue while retaining fixed-capacity historical banks for recurrent states. An exact active-set-size-matched recent control tests whether any difference is explained only by the number of active residuals, and an exact active-set-size-matched regime-agnostic replay control tests whether generic access to older history is sufficient. We further examine whether the benefit of retaining historical residuals is concentrated near the beginning of recurrent stress episodes.

Recent-only FIFO calibration remains close to its aggregate coverage target but undercovers substantially on high-volatility dates in both the ETF development panel and the fixed stock panel. State-structured historical access yields a lower stress-period interval score than recent-only memory and the two matched controls, while the contrast is largest near stress re-entry and becomes smaller later in the episode. Our analysis demonstrates how the composition of bounded calibration memory influences state-specific forecasting reliability. By comparing recent-only calibration, state-structured replay, and matched historical controls, we assess the importance of retaining residual information from recurrent market states.

The remainder of the paper is organized as follows. Section~\ref{sec:setup} defines the common forecasting model and calibration rule. Section~\ref{sec:diagnostic} describes the controlled information-set comparisons. Section~\ref{sec:design} gives the empirical design and time-series inference. Section~\ref{sec:results} reports the main findings and stress re-entry analysis. Section~\ref{sec:discussion} discusses implications and limitations.

\section{Forecasting Setup and Calibration Memory}
\label{sec:setup}

\subsection{Common forecasting model and calibration rule}

For asset $i$ and target date $t$, the response is the next-trading-day log high--low range
\[
Y_{i,t}=\log\left(\frac{H_{i,t}}{L_{i,t}}\right).
\]
Range-based volatility measures are classical summaries of intraday price variation \citep{parkinson1980range,chou2005carr}. The point forecast is the standard heterogeneous autoregressive (HAR) specification \citep{corsi2009har}, fitted sequentially using only information available before the target:
\[
\widehat Y_{i,t}=\left[\widehat\beta_{0,i,t}+\widehat\beta_{1,i,t}Y_{i,t-1}
+\widehat\beta_{5,i,t}\overline Y_{i,t-1}^{(5)}
+\widehat\beta_{22,i,t}\overline Y_{i,t-1}^{(22)}\right]_+,
\]
where $\overline Y_{i,t-1}^{(d)}$ is the mean of the previous $d$ observed ranges and $[x]_+=\max\{x,0\}$. To isolate calibration-memory behavior, this forecasting backbone is held fixed in the main analysis; a lagged-VIX extension in the Appendix tests whether the main pattern depends on omitting VIX from the point forecast.

After the target is observed, the calibration score is the absolute forecast error
\[
R_{i,t}=|Y_{i,t}-\widehat Y_{i,t}|.
\]
All main designs use the same finite-sample conformal cutoff rule, the same adaptive conformal inference update, and the same sequential order: form the interval, observe the target, update the controller, and then insert the new residual. Only the stored calibration history and the records activated at a query differ across the information-set comparisons. Exact cutoff and controller formulas, together with additional weighting designs, are given in the Appendix.

\subsection{Bounded calibration memory}

The reference design is a FIFO buffer that stores only the most recent $B$ residual records. When a new record arrives after the buffer is full, the oldest record is removed. This rule is simple and natural under a fixed storage budget, but it represents history entirely through recency. A state that has not appeared for sufficiently long can therefore disappear from the calibration memory even when it later recurs.

Storage determines which historical residuals remain available, retrieval determines which stored residuals are active at the current query, and weighting determines how active residuals contribute to the cutoff. Alternative state-aware organization and weighting are examined in the Appendix.

\section{Controlled Information-Set Comparisons}
\label{sec:diagnostic}

Table~\ref{tab:designs} summarizes the calibration information sets used in the main comparisons and their diagnostic roles. The \UB is used only as an information benchmark.

\begin{table}[H]
\centering
\begin{threeparttable}
\caption{Calibration information sets used in the main comparisons. The unbounded reference is not a fixed-budget competitor.}
\label{tab:designs}
\small
\begin{tabular}{p{0.17\linewidth}p{0.48\linewidth}p{0.27\linewidth}}
\toprule
Design & Active calibration evidence & Diagnostic purpose\\
\midrule
FIFO & Most recent $B$ residuals & Recent-only reference\\
UB reference & Same recent queue as Rare replay plus all older residuals from the current state & Assess the contribution of historical residuals discarded by FIFO\\
Recent-$n$ & Most recent $n$ residuals; $n$ matches Rare replay's active size & Examine whether differences arise from active-set size\\
Hybrid-$n$ & Recent queue plus a size-matched sample from a global historical reservoir & Evaluate whether regime-agnostic historical access explains the observed improvement\\
Rare replay & Recent queue plus records from a bounded current-state bank & Assess state-structured retention and retrieval of recurrent-state history\\
\bottomrule
\end{tabular}
\end{threeparttable}
\end{table}

The structured-replay design keeps recent residuals active across states and uses state-specific reservoirs for older observations, activating only the bank associated with the current pre-target state. The replay allocation assigns greater historical capacity to less frequent validation states. Both matched controls use exactly the realized active size of Rare replay; Hybrid-$n$ uses the same recent--historical split but a regime-agnostic reservoir. Allocation details and additional calibration designs are given in the Appendix.

\section{Empirical Design}
\label{sec:design}

\subsection{Data, states, and chronology}

The analysis uses daily open--high--low--close (OHLC) data from Yahoo Finance with automatic price adjustment disabled. The development panel contains 11 exchange-traded funds: SPY, QQQ, IWM, XLK, XLF, XLE, XLV, XLY, XLP, TLT, and GLD. The second panel is a prespecified set of 50 large, liquid U.S.-listed equities with complete OHLC histories over validation and evaluation. We refer to the two panels as ETF11 and Stock50. Stock50 is neither probability sampled nor survivorship-free and is not intended to represent the U.S. equity market; it is used to assess whether the analysis transfers to a broader cross-sectional panel without additional parameter selection.

The raw sample spans 2005--2025. The years 2005--2008 provide warm-up, 2009--2011 form the validation period, and the evaluation period runs from January 3, 2012 through December 31, 2025. Both panels contain 3,520 evaluation dates per asset.

The Cboe Volatility Index (VIX) is shifted by one observation on its own trading calendar before being joined to each asset's target dates, so the regime is observable before the target. We define
\[
G_t=\begin{cases}
L,& \mathrm{VIX}_{t-1}<20,\\
M,& 20\le \mathrm{VIX}_{t-1}<30,\\
H,& \mathrm{VIX}_{t-1}\ge 30.
\end{cases}
\]
The thresholds are specified before evaluation and are not selected using evaluation outcomes. VIX enters memory organization and state-specific evaluation, but not the HAR point forecast. Missing merged rows are removed rather than forward-filled. Each calibration design is replayed sequentially over the pre-evaluation residual stream so that its memory state carries into evaluation. At the start of evaluation, the adaptive level is reset to 0.10 for every design and randomized seed; subsequent controller paths evolve separately.

\subsection{Evaluation and time-series inference}

The main empirical target is the FIFO high-minus-non-high coverage gap on the common evaluation calendar. For comparisons between calibration information sets, we use the interval score at nominal miscoverage $0.10$ \citep{gneiting2007scoring}; lower values are better. Interval score penalizes both width and noncoverage, while coverage and width are reported alongside it for interpretation. The evaluation calendar contains 156 high-VIX dates.

The main memory budget is $B=250$; budget sensitivity is reported in the Appendix. State frequencies used by the structured-memory designs are estimated on the ETF validation period, and the same configuration is then used for Stock50. Randomized retention designs are averaged over 20 prespecified reservoir seeds.

Pointwise 95\% confidence intervals use 5,000 paired circular block-bootstrap draws over the shared 3,520-date calendar. The main block length is 20 trading days (approximately one trading month), with 10- and 40-day sensitivity analyses. State labels travel with resampled dates, and state-specific means and coverage gaps are recalculated within each draw. Metrics are first computed within asset and then averaged equally across assets. Because both panels share the same market calendar, the 61 securities are not treated as independent temporal replications.

For the re-entry analysis, a stress entry is a high-state date preceded by 20 non-high-state prediction origins. Later entries are classified as recurrent, and a strict FIFO-eviction entry has no high-state observation among the preceding $B$ origins. Within recurrent episodes, we compare the first five high-state dates with later high-state dates and repeat the pooled \Rare-minus-FIFO calculation after omitting each episode in turn. These episode analyses are descriptive timing analyses rather than independent inferential replications.

\section{Empirical Results}
\label{sec:results}

\subsection{Aggregate calibration masks stress-state failure}
\label{subsec:gap}

The central empirical result is the divergence between aggregate and state-conditional coverage under FIFO. Overall coverage is 0.9003 in ETF11 and 0.9002 in Stock50, essentially equal to the nominal 0.90 target. On the 156 high-VIX dates, however, coverage falls to 0.7576 and 0.7763; on the remaining dates it is 0.9070 and 0.9059. Thus the aggregate result conceals high-minus-non-high coverage gaps of $-0.1494$ and $-0.1297$.

\begin{figure}[H]
\centering
\includegraphics[width=0.84\textwidth]{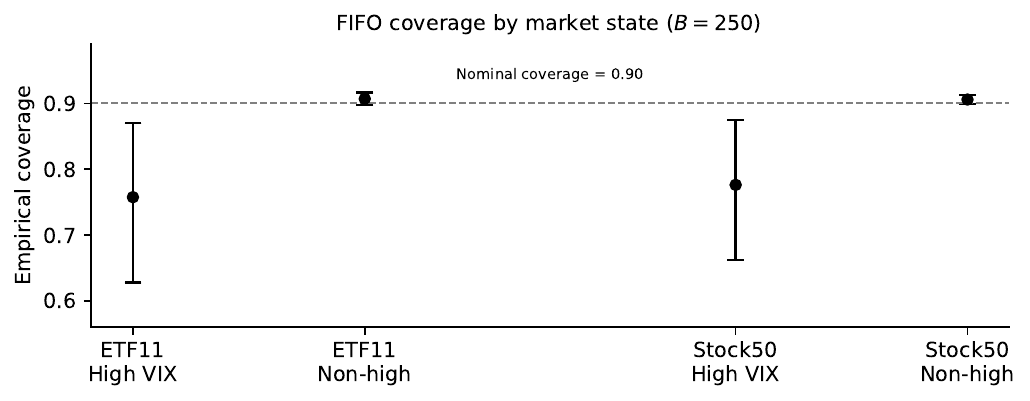}
\caption{FIFO empirical coverage at $B=250$. Points show asset-averaged coverage and vertical bars show pointwise 95\% paired circular block-bootstrap intervals with 20-day blocks. The dashed line is the nominal 0.90 target.}
\label{fig:coverage}
\end{figure}

Figure~\ref{fig:coverage} displays the state-specific coverage estimates and their block-bootstrap intervals. The high-minus-non-high gap has a 95\% paired block-bootstrap interval of $[-0.2778,-0.0376]$ in ETF11 and $[-0.2440,-0.0318]$ in Stock50. Relative to the nominal 0.90 target, the corresponding high-state gaps are also separated from zero. The same recent-only system that appears satisfactory in aggregate therefore has a materially different reliability profile under observable stress.

\subsection{Historical availability beyond simpler explanations}
\label{subsec:availability}

Table~\ref{tab:availability} reports the main information-set comparisons. The \UB has a lower observed stress-period interval score than FIFO in both panels and serves as a direct information reference for history discarded by recent-only memory. Among bounded designs, \Rare also has a lower stress-period score than FIFO. The two matched controls address simpler explanations: \RecentN matches the active-set size using only recent residuals, whereas \HybridN additionally restores size-matched regime-agnostic historical access. Neither reproduces the stress-period performance of state-structured replay.

\begin{table}[H]
\centering
\begin{threeparttable}
\caption{Historical-availability comparisons at $B=250$. Interval score is evaluated on high-VIX dates; lower is better.}
\label{tab:availability}
\small
\begin{tabular}{lcccccc}
\toprule
& \multicolumn{3}{c}{ETF11} & \multicolumn{3}{c}{Stock50}\\
\cmidrule(lr){2-4}\cmidrule(lr){5-7}
Design & Coverage & Width & IS & Coverage & Width & IS\\
\midrule
\FIFO & 0.7576 & 0.0261 & 0.0872 & 0.7763 & 0.0367 & 0.1058\\
\RecentN & 0.7797 & 0.0287 & 0.0839 & 0.7976 & 0.0400 & 0.1028\\
\HybridN & 0.7714 & 0.0274 & 0.0855 & 0.7924 & 0.0388 & 0.1033\\
\Rare & 0.8478 & 0.0345 & 0.0736 & 0.8619 & 0.0473 & 0.0925\\
\UB & 0.9009 & 0.0423 & 0.0688 & 0.9092 & 0.0571 & 0.0878\\
\bottomrule
\end{tabular}
\begin{tablenotes}[flushleft]\footnotesize
\item Notes: The unbounded current-state reference is an information benchmark rather than a bounded competitor. All bounded comparisons respect the same storage cap; \RecentN and \HybridN exactly match the realized active size of \Rare.
\end{tablenotes}
\end{threeparttable}
\end{table}

Table~\ref{tab:contrasts} shows that \Rare achieves lower high-VIX interval scores than FIFO and both active-size-matched controls in ETF11 and Stock50, with the reported pointwise confidence intervals excluding zero. The \UB contrast is favorable in both panels, with its interval excluding zero in ETF11 but not Stock50.

\begin{table}[H]
\centering
\caption{Paired high-VIX interval-score comparisons at $B=250$. Estimates are first design minus second; negative values favor the first. Pointwise 95\% intervals use the 20-day paired circular block bootstrap, are not adjusted for multiplicity, and share the same market calendar across panels.}
\label{tab:contrasts}
\scriptsize
\setlength{\tabcolsep}{4pt}
\begin{tabular}{lcc}
\toprule
Comparison & ETF11 [95\% CI] & Stock50 [95\% CI]\\
\midrule
Unbounded current-state $-$ \FIFO & -0.0184 [-0.0408, -0.0002] & -0.0180 [-0.0420, 0.0021]\\
\Rare $-$ \FIFO & -0.0136 [-0.0275, -0.0018] & -0.0132 [-0.0286, -0.0009]\\
\Rare $-$ \RecentN & -0.0102 [-0.0213, -0.0019] & -0.0103 [-0.0212, -0.0014]\\
\Rare $-$ \HybridN & -0.0119 [-0.0242, -0.0017] & -0.0108 [-0.0224, -0.0013]\\
\bottomrule
\end{tabular}
\end{table}

Because neither matched control attains the Rare replay interval score, the difference is not explained by active-set size or regime-agnostic access to older residuals. Because structured replay changes storage and retrieval together, these comparisons do not separately identify the effects of storage and retrieval.

\subsection{Stress re-entry and FIFO refill}
\label{subsec:reentry}

We further examine whether the benefit of retaining historical residuals is concentrated near the beginning of recurrent stress episodes. The evaluation calendar contains 13 stress entries, 12 of which are recurrent, and three recurrent entries satisfy the strict FIFO-eviction criterion. At recurrent entries, high-state residuals account for 9.57\% of FIFO's active set and 34.44\% of the \Rare active set. The historical replay component has 48 active residuals at these entries, with a median age of approximately 2,914 prediction origins, so the state-aware design is not merely drawing from the recent edge of the buffer.

Table~\ref{tab:reentry} shows the re-entry pattern. The Rare-minus-FIFO interval-score difference is more negative during the first five high-state dates than later in the episode in both panels and is most negative at the strict FIFO-eviction entries. As FIFO accumulates new high-state residuals, the difference moves toward zero. This is the temporal pattern expected if historical availability matters most when a recurrent state first returns and less after recent memory has refilled.

\begin{table}[H]
\centering
\caption{Stress re-entry analysis at $B=250$. Interval-score differences are \Rare minus FIFO; negative values favor \Rare. The strict-eviction row contains only three dates and is descriptive.}
\label{tab:reentry}
\small
\begin{tabular}{lcccc}
\toprule
Window & FIFO high-state share & Rare high-state share & ETF11 $\Delta$IS & Stock50 $\Delta$IS\\
\midrule
Recurrent entries & 0.0957 & 0.3444 & -0.0167 & -0.0199\\
First five high-state dates & 0.1037 & 0.3513 & -0.0232 & -0.0202\\
Later high-state dates & 0.1471 & 0.4814 & -0.0052 & -0.0081\\
Strict FIFO eviction & 0.0000 & 0.2775 & -0.0386 & -0.0505\\
\bottomrule
\end{tabular}
\end{table}

The episode pattern is not uniform: \Rare has lower episode-specific interval score than FIFO in 10 of 13 ETF11 episodes and 9 of 13 Stock50 episodes. We therefore treat this analysis as a timing diagnostic rather than episode-level replication. A leave-one-episode-out analysis nevertheless leaves the pooled stress-period contrast negative after omitting any one episode in either panel, showing that the pooled result is not generated by a single stress episode.

Additional robustness and sensitivity analyses are reported in the Appendix.

\section{Discussion}
\label{sec:discussion}

Aggregate calibration is an incomplete reliability criterion when forecast uncertainty is evaluated under recurrent states. In the main analysis, recent-only FIFO calibration meets its long-run coverage target while providing materially lower coverage during high-volatility periods. The relevant question is therefore not only whether intervals are calibrated on average, but whether the calibration information set remains informative when an economically important state returns.

The controlled comparisons separate historical information availability from two simpler explanations. Exact active-size matching and regime-agnostic historical access do not reproduce the state-structured result, and the contrast is strongest before FIFO has rebuilt stress history. In the Appendix, predictor and controller robustness analyses preserve the Rare-versus-FIFO pattern, while a historical-age control does not reproduce the state-structured result. Historical information availability is therefore a distinct design dimension in this system; because replay changes storage and retrieval jointly, their separate causal contributions are not identified.

For practice, bounded calibrators should be evaluated along two axes: state-specific reliability and memory composition. Increasing a recent window can delay eviction but does not control which recurrent states remain represented; similarly, reweighting can act only on records that survived storage. This distinction matters whenever the forecasting problem contains observable recurrent states and a binding calibration-memory constraint.

The evidence remains conditional on HAR-based range forecasts, two fixed U.S.-listed asset panels, and the observable state definitions used here. Stock50 provides a cross-sectional transfer analysis rather than an independent temporal replication because both panels share the same market calendar, and 13 stress episodes are too few for episode-level inference. Future work can examine other forecasting backbones, markets, and learned state representations, or use designs that vary storage and retrieval separately. Calibration memory is part of the forecasting design because it determines which historical uncertainty information exists when a state returns.

\section*{AI Use Statement}
OpenAI ChatGPT was used for language polishing and coding assistance. The authors reviewed the resulting edits and code and take full responsibility for the final manuscript and analyses.

\section*{Conflict of Interest}
The authors declare no conflicts of interest.

\section*{Funding}
This research received no specific grant from any funding agency in the public, commercial, or not-for-profit sectors.

\section*{Data Availability Statement}
The code and derived analysis files supporting the reported results are available from the corresponding author on reasonable request. Market data were obtained from Yahoo Finance and Cboe and remain subject to the providers' terms of use.

\section*{Ethics Statement}
Not applicable; the study uses historical market data and does not involve human participants.

\printbibliography

\clearpage
\appendix
\section{Data and common forecasting implementation}

All analyses use the same Yahoo Finance market data as the main analysis, with automatic price adjustment disabled. ETF11 contains SPY, QQQ, IWM, XLK, XLF, XLE, XLV, XLY, XLP, TLT, and GLD. Stock50 contains AAPL, MSFT, AMZN, GOOGL, NVDA, JPM, BAC, GS, MS, XOM, CVX, JNJ, PFE, MRK, WMT, COST, HD, MCD, KO, PEP, PG, NKE, DIS, ADBE, CRM, CSCO, INTC, ORCL, IBM, QCOM, ABT, TMO, LLY, UNH, MDT, AMGN, HON, CAT, DE, BA, GE, MMM, UPS, LOW, SBUX, TGT, CL, MO, SLB, and TXN. Stock50 was prespecified before its transfer analysis; it is neither probability sampled nor survivorship-free.

The one-sided HAR coefficients are re-estimated by ordinary least squares at every target date using at most the 756 most recent valid origin--response pairs, with an intercept; forecast generation begins once the raw range series reaches index 252. The point forecast is truncated at zero. The VIX series is shifted by one observation on its own trading calendar before being joined to asset target dates. Missing merged rows are removed and are not forward-filled.

\section{Calibration implementation}

For asset $i$ and target date $t$, let $\mathcal A_{i,t}$ denote the active calibration records and let $a_{i,s,t}\ge0$ denote their query-time weights. With $W_{i,t}=\sum_{s\in\mathcal A_{i,t}}a_{i,s,t}$, the finite-sample cutoff used throughout the main analysis is
\[
\widehat q_{i,t}=\inf\left\{q:\sum_{s\in\mathcal A_{i,t}}a_{i,s,t}\mathbf 1\{R_{i,s}\le q\}
\ge (1-\alpha_{i,t})(W_{i,t}+1)\right\}.
\]
If the required mass exceeds $W_{i,t}$, the largest active score is used. The interval is
\[
C_{i,t}=\left[\max\{0,\widehat Y_{i,t}-\widehat q_{i,t}\},\ \widehat Y_{i,t}+\widehat q_{i,t}\right].
\]
In the main analysis, after observing the target, $V_{i,t}=\mathbf 1\{Y_{i,t}\notin C_{i,t}\}$ and
\[
\alpha_{i,t+1}=\Pi_{[0.001,0.30]}\left[\alpha_{i,t}+0.005\{0.10-V_{i,t}\}\right].
\]
At the start of evaluation, $\alpha_{i,t_0}=0.10$ for every design and randomized seed; subsequent controller paths evolve separately. The sequential order is always interval formation, target observation, controller update, and residual insertion.

The structured-replay design divides the budget into $B_{\mathrm{rec}}=\lfloor B/2\rfloor$ recent slots and $B_{\mathrm{rep}}=B-B_{\mathrm{rec}}$ historical slots. Let $\widehat p_g$ be the validation-period frequency of state $g\in\{L,M,H\}$. For allocation exponent $\eta$, the unrounded capacity of bank $g$ is
\[
\widetilde B_g(\eta)=B_{\mathrm{rep}}\frac{\widehat p_g^{\eta}}{\sum_{h\in\{L,M,H\}}\widehat p_h^{\eta}},
\]
followed by largest-remainder rounding. Equal, frequency-proportional, and inverse-frequency allocations use $\eta=0,1,-1$, respectively; Rare uses the inverse-frequency allocation selected before evaluation. At $B=250$, its recent queue has 125 slots and its $(L,M,H)$ historical-bank capacities are $(46,31,48)$. Each bank uses standard reservoir sampling. An evicted residual is routed to the bank associated with its stored pre-target VIX state, and the active set at a query is the recent queue together with the bank matching the current state.

Recent-$n$ uses only the most recent residuals, with $n$ set to the realized active-set size of Rare at each asset and date. Hybrid-$n$ uses the same recent--historical split and a global reservoir with capacity $B_{\mathrm{rep}}$. At each asset--date--seed it activates a regime-agnostic priority subsample whose size exactly matches Rare's active replay component. Thus both matched controls reproduce Rare's active-set size without using its state-structured historical composition.

The bounded Mondrian reconstruction stores observations separately by VIX state under the same total budget. The proportional version allocates state capacities according to validation-period frequencies and uses only the current-state buffer at a query; its adaptive controller is state specific.

The weighting designs all store the most recent $B$ residuals. Time-weighted calibration uses
\[
a^{\mathrm{TWC}}_{i,s,t}=2^{-d_{i,s,t}/h_{1/2}},
\]
where $d_{i,s,t}$ is the age in prediction origins. Regime-aware weighting uses
\[
a^{\mathrm{RWC}}_{i,s,t}=2^{-d_{i,s,t}/h_{1/2}}
\exp\{-\gamma|c(G_s)-c(G_t)|\},
\]
with $c(L)=0$, $c(M)=1$, and $c(H)=2$. RWC-MM rescales the regime-aware weights to match the contemporaneous total mass of TWC. TWC is selected over $h_{1/2}\in\{0.25B,0.50B,B,\infty\}$ with $\gamma=0$. RWC-adapted is selected over the same half-life grid crossed with $\gamma\in\{0,0.5,1,2\}$. Selection is performed on ETF11 validation data. Candidates with overall validation coverage in $[0.89,0.91]$ are ranked first by high-VIX validation interval score, then overall validation interval score, smaller $\gamma$, and shorter half-life. If none meets the coverage constraint, candidates are ranked first by absolute distance from 0.90 overall coverage, followed by the same tie-breaks. At $B=250$, TWC selects $h_{1/2}=62.5$ and RWC-adapted selects $(h_{1/2},\gamma)=(62.5,1)$; the same parameter values are then used for Stock50. RWC-MM is a mass-matched comparison based on the selected regime-aware profile and is not tuned separately.

\section{Calendar-block uncertainty for the core coverage gap}

Table~\ref{tab:gap-sens} reports the FIFO high-minus-non-high coverage gap under 10-, 20-, and 40-trading-day circular block lengths. State labels are resampled with dates and condition-specific coverage is recomputed within every bootstrap draw. The gap remains negative and its pointwise 95\% interval excludes zero in both panels at all three block lengths.

\begin{table}[h]
\centering
\caption{FIFO high-VIX minus non-high-VIX coverage gap.}
\label{tab:gap-sens}
\small
\begin{tabular}{lccc}
\toprule
Panel & Block length & Estimate & 95\% CI\\
\midrule
ETF11 & 10 & -0.1494 & [-0.2610, -0.0495]\\
ETF11 & 20 & -0.1494 & [-0.2778, -0.0376]\\
ETF11 & 40 & -0.1494 & [-0.2865, -0.0352]\\
Stock50 & 10 & -0.1297 & [-0.2284, -0.0430]\\
Stock50 & 20 & -0.1297 & [-0.2440, -0.0318]\\
Stock50 & 40 & -0.1297 & [-0.2496, -0.0276]\\
\bottomrule
\end{tabular}
\end{table}

For reference, the 20-day intervals for FIFO high-state coverage are [0.6280, 0.8698] in ETF11 and [0.6617, 0.8752] in Stock50. The corresponding non-high-state intervals are [0.8972, 0.9166] and [0.8987, 0.9132].

\section{Predictor and controller robustness}
\label{sec:pred-controller}

The main analysis uses the HAR point predictor and adaptive conformal controller. We assess sensitivity to these choices by augmenting the rolling HAR with one strictly pre-target regressor, $\log(\mathrm{VIX}_{t-1})$, using the same rolling window of up to 756 origin--response pairs and no additional tuning. A second analysis sets $\alpha_t=0.10$ on every date and therefore removes controller feedback. The two modifications are crossed at $B=250$, with Rare replay averaged over the same 20 reservoir seeds used in the main analysis.

\begin{table}[h]
\centering
\caption{Predictor and controller robustness at $B=250$. The final column is Rare minus FIFO high-VIX interval score; negative values favor Rare.}
\label{tab:pred-controller}
\scriptsize
\begin{tabular}{lllrrrr}
\toprule
Panel & Predictor & Controller & FIFO overall cov. & FIFO high cov. & Rare high cov. & $\Delta$IS [95\% CI]\\
\midrule
ETF11 & HAR & ACI & 0.9003 & 0.7576 & 0.8478 & -0.01361 [-0.02754,-0.00197]\\
ETF11 & HAR & Fixed $\alpha$ & 0.9033 & 0.6987 & 0.8297 & -0.01723 [-0.03312,-0.00305]\\
ETF11 & HAR-VIX & ACI & 0.9004 & 0.7459 & 0.8532 & -0.01113 [-0.02208,-0.00162]\\
ETF11 & HAR-VIX & Fixed $\alpha$ & 0.9013 & 0.6795 & 0.8420 & -0.01473 [-0.02741,-0.00334]\\
\addlinespace
Stock50 & HAR & ACI & 0.9002 & 0.7763 & 0.8619 & -0.01321 [-0.02811,-0.00094]\\
Stock50 & HAR & Fixed $\alpha$ & 0.9007 & 0.7265 & 0.8414 & -0.01702 [-0.03393,-0.00201]\\
Stock50 & HAR-VIX & ACI & 0.9003 & 0.7636 & 0.8638 & -0.01200 [-0.02510,-0.00124]\\
Stock50 & HAR-VIX & Fixed $\alpha$ & 0.9012 & 0.7106 & 0.8514 & -0.01532 [-0.02988,-0.00246]\\
\bottomrule
\end{tabular}
\end{table}

For the HAR+ACI baseline, the point estimates match the main analysis; the bootstrap intervals are recomputed for this sensitivity analysis, so their endpoints can differ slightly from those reported in the main text.

Every Rare-minus-FIFO contrast is negative and every reported calendar-block interval excludes zero. Rare has a lower interval score in all 11 ETF assets and all 50 stock assets in each predictor--controller cell. The FIFO high-minus-non-high coverage gap also remains large after lagged VIX is added to the point forecast. Thus the main pattern is not removed by giving the point predictor direct pre-target VIX information or by eliminating adaptive-controller feedback.

\section{Historical-age control}
\label{sec:age}

The Hybrid-$n$ control restores regime-agnostic historical access at the same active size as Rare, but its replay observations are younger on average. To test whether historical age alone explains the difference, we add an Oldest-$n$ control under the HAR model with fixed $\alpha=0.10$. Oldest-$n$ uses the same bounded global reservoir as Hybrid-$n$ and activates the same number of historical records as Rare at each query, but deliberately chooses the oldest available records without using regime labels.

\begin{table}[h]
\centering
\caption{Historical-age control at $B=250$ under fixed $\alpha=0.10$.}
\label{tab:age}
\small
\begin{tabular}{llrrr}
\toprule
Panel & Design & Replay median age & High-VIX coverage & High-VIX IS\\
\midrule
ETF11 & Hybrid-$n$ & 1962 & 0.7424 & 0.08745\\
ETF11 & Oldest-$n$ & 3122 & 0.7606 & 0.08509\\
ETF11 & Rare replay & 2784 & 0.8297 & 0.07554\\
\addlinespace
Stock50 & Hybrid-$n$ & 1959 & 0.7621 & 0.10683\\
Stock50 & Oldest-$n$ & 3109 & 0.7792 & 0.10417\\
Stock50 & Rare replay & 2785 & 0.8414 & 0.09535\\
\bottomrule
\end{tabular}
\end{table}

Oldest-$n$ retrieves observations that are older than those used by Rare in both panels, yet Rare retains the lower stress-period interval score. Rare minus Oldest-$n$ is -0.00956 in ETF11 (95\% CI [-0.01978,-0.00067]) and -0.00882 in Stock50 ([-0.01875,-0.00017]); Rare has a lower interval score in all 11 ETF assets and all 50 stock assets, respectively. Historical age by itself therefore does not reproduce the state-structured result.

\section{Alternative pre-target stress definition}
\label{sec:altstress}

The VIX state is used for memory organization and for the main state-specific evaluation. As a separate sensitivity analysis, stress is redefined from SPY trailing 22-trading-day close-to-close realized volatility. The measure is shifted by one SPY trading day so that it is available before the target, and the stress threshold is fixed at the 95th percentile of the 2009--2011 validation period before being applied unchanged to 2012--2025. This definition yields 56 evaluation stress dates, 43 of which overlap the 156 VIX-high dates.

\begin{table}[H]
\centering
\caption{HAR+ACI results under the alternative pre-target SPY realized-volatility stress definition.}
\label{tab:altstress}
\small
\begin{tabular}{lrrr}
\toprule
Panel & FIFO stress coverage & Rare stress coverage & Rare $-$ FIFO stress IS [95\% CI]\\
\midrule
ETF11 & 0.7305 & 0.8396 & -0.01875 [-0.0531,0.0033]\\
Stock50 & 0.7600 & 0.8500 & -0.01902 [-0.0594,0.0035]\\
\bottomrule
\end{tabular}
\end{table}

Rare has a lower stress-period interval score in all 11 ETF assets and all 50 stock assets. The calendar-block intervals cross zero, however, and the alternative subgroup is small and substantially overlaps the VIX-high dates. The result therefore assesses sensitivity to the exact VIX-based subgroup definition rather than providing an independent replication.

\section{Policy summaries at B = 250}

Table~\ref{tab:full-primary} reports the high-VIX coverage, width, and interval score for the main historical-availability designs together with additional state-aware and weighting comparisons. The main text focuses on the historical-availability comparisons.

\begin{table}[h]
\centering
\caption{High-VIX performance at $B=250$. Lower interval score is better.}
\label{tab:full-primary}
\small
\begin{tabular}{lcccccc}
\toprule
& \multicolumn{3}{c}{ETF11} & \multicolumn{3}{c}{Stock50}\\
\cmidrule(lr){2-4}\cmidrule(lr){5-7}
Design & Coverage & Width & IS & Coverage & Width & IS\\
\midrule
FIFO & 0.7576 & 0.0261 & 0.0872 & 0.7763 & 0.0367 & 0.1058\\
Recent-$n$ & 0.7797 & 0.0287 & 0.0839 & 0.7976 & 0.0400 & 0.1028\\
Hybrid-$n$ & 0.7714 & 0.0274 & 0.0855 & 0.7924 & 0.0388 & 0.1033\\
Rare replay & 0.8478 & 0.0345 & 0.0736 & 0.8619 & 0.0473 & 0.0925\\
Mondrian proportional & 0.8858 & 0.0400 & 0.0709 & 0.8824 & 0.0550 & 0.0941\\
TWC & 0.7949 & 0.0302 & 0.0800 & 0.8159 & 0.0418 & 0.0983\\
RWC-MM & 0.8671 & 0.0391 & 0.0706 & 0.8742 & 0.0525 & 0.0898\\
RWC-adapted & 0.9009 & 0.0428 & 0.0636 & 0.9023 & 0.0568 & 0.0851\\
Unbounded current-state & 0.9009 & 0.0423 & 0.0688 & 0.9092 & 0.0571 & 0.0878\\
\bottomrule
\end{tabular}
\end{table}

Across these designs, overall coverage lies between 0.9001 and 0.9011 in ETF11 and between 0.8987 and 0.9008 in Stock50.

\section{State-specific policy summaries}

Table~\ref{tab:state} reports low-, medium-, and high-VIX performance for FIFO, Rare replay, and the proportional Mondrian reconstruction. State counts are common across assets: 2,634 low-VIX dates, 730 medium-VIX dates, and 156 high-VIX dates.

\begin{longtable}{llrrrr}
\caption{State-specific performance at $B=250$.}\label{tab:state}\\
\toprule
Panel & Design & State & Coverage & Interval score & Width\\
\midrule
\endfirsthead
\toprule
Panel & Design & State & Coverage & Interval score & Width\\
\midrule
\endhead
ETF11 & FIFO & L & 0.9158 & 0.02107 & 0.01516\\
ETF11 & FIFO & M & 0.8751 & 0.03703 & 0.02212\\
ETF11 & FIFO & H & 0.7576 & 0.08724 & 0.02606\\
ETF11 & Rare replay & L & 0.9100 & 0.02084 & 0.01453\\
ETF11 & Rare replay & M & 0.8788 & 0.03642 & 0.02195\\
ETF11 & Rare replay & H & 0.8478 & 0.07363 & 0.03453\\
ETF11 & Mondrian proportional & L & 0.9008 & 0.02064 & 0.01392\\
ETF11 & Mondrian proportional & M & 0.9022 & 0.03508 & 0.02338\\
ETF11 & Mondrian proportional & H & 0.8858 & 0.07091 & 0.03997\\
\addlinespace
Stock50 & FIFO & L & 0.9107 & 0.03353 & 0.02223\\
Stock50 & FIFO & M & 0.8887 & 0.04973 & 0.03066\\
Stock50 & FIFO & H & 0.7763 & 0.10575 & 0.03675\\
Stock50 & Rare replay & L & 0.9056 & 0.03333 & 0.02152\\
Stock50 & Rare replay & M & 0.8886 & 0.04905 & 0.03010\\
Stock50 & Rare replay & H & 0.8619 & 0.09254 & 0.04731\\
Stock50 & Mondrian proportional & L & 0.9006 & 0.03340 & 0.02136\\
Stock50 & Mondrian proportional & M & 0.9007 & 0.04790 & 0.03084\\
Stock50 & Mondrian proportional & H & 0.8824 & 0.09411 & 0.05498\\
\bottomrule
\end{longtable}

\section{Alternative state-aware and weighting comparisons}

Table~\ref{tab:alt-sens} reports paired high-VIX interval-score contrasts for the proportional Mondrian reconstruction and query-time weighting branch.

\begin{table}[h]
\centering
\caption{High-VIX interval-score contrasts. Estimates are first design minus second.}
\label{tab:alt-sens}
\scriptsize
\begin{tabular}{@{}llrrrr@{}}
\toprule
Panel & Contrast & Estimate & 10-day block & 20-day block & 40-day block\\
\midrule
ETF11 & Rare $-$ FIFO & -0.01361 & [-0.02603,-0.00293] & [-0.02754,-0.00196] & [-0.02828,-0.00161]\\
ETF11 & Mondrian $-$ FIFO & -0.01632 & [-0.03537,-0.00011] & [-0.03764,0.00154] & [-0.03859,0.00203]\\
ETF11 & Rare $-$ Mondrian & 0.00272 & [-0.00307,0.00969] & [-0.00367,0.01035] & [-0.00410,0.01054]\\
ETF11 & RWC-MM $-$ TWC & -0.0093 & --- & [-0.0224,0.0011] & ---\\
ETF11 & RWC-adapted $-$ RWC-MM & -0.0070 & --- & [-0.0162,-0.0011] & ---\\
ETF11 & RWC-adapted $-$ Rare & -0.0100 & --- & [-0.0253,0.0011] & ---\\
\addlinespace
Stock50 & Rare $-$ FIFO & -0.01321 & [-0.02632,-0.00221] & [-0.02811,-0.00093] & [-0.02860,-0.00088]\\
Stock50 & Mondrian $-$ FIFO & -0.01164 & [-0.03024,0.00344] & [-0.03280,0.00493] & [-0.03270,0.00534]\\
Stock50 & Rare $-$ Mondrian & -0.00157 & [-0.00629,0.00465] & [-0.00672,0.00498] & [-0.00688,0.00483]\\
Stock50 & RWC-MM $-$ TWC & -0.0085 & --- & [-0.0224,0.0023] & ---\\
Stock50 & RWC-adapted $-$ RWC-MM & -0.0047 & --- & [-0.0109,-0.0004] & ---\\
Stock50 & RWC-adapted $-$ Rare & -0.0074 & --- & [-0.0205,0.0029] & ---\\
\bottomrule
\end{tabular}
\end{table}

Mondrian has a lower point estimate than FIFO in both panels, but the 20-day block-bootstrap intervals include zero. Rare versus Mondrian changes sign across panels, and all reported block-bootstrap intervals include zero. The 20-day interval for RWC-MM versus TWC includes zero, whereas the interval for RWC-adapted versus RWC-MM does not; the latter comparison also changes total calibration mass and therefore does not isolate relative state weighting alone. These results do not identify a uniformly superior state-aware implementation.

\section{Stress-entry concentration and additional robustness}

The evaluation calendar contains 13 high-state entries, 12 of which are recurrent and three of which meet the strict FIFO-eviction definition. Rare replay has lower episode-specific mean interval score than FIFO in 10/13 ETF11 episodes and 9/13 Stock50 episodes. The pooled contrast remains negative after omitting any one episode: for ETF11, the leave-one-episode-out pooled effect ranges from -0.01539 to -0.00994; for Stock50, from -0.01471 to -0.00991. The 13 common episodes are not treated as independent inferential units.

Additional sensitivity analyses are as follows.
\begin{itemize}
\item At $B=500$, Rare replay remains below FIFO in stress-period interval score in ETF11 (-0.01365, 95\% CI [-0.02778,-0.00207]) and Stock50 (-0.01266, [-0.02481,-0.00154]). Rare replay also remains below the exact active-size recent control in both panels.
\item Under the 20/30 VIX partition used in the main analysis, equal, frequency-proportional, and inverse-frequency structured replay all have lower observed stress-period interval scores than the recent-only controls. Under an alternative 18/28 partition in ETF11, the Rare-minus-proportional difference is -0.00011 with a 95\% CI of [-0.00085,0.00051].
\item The Rare-minus-FIFO interval-score difference remains negative for all 20 reservoir seeds and for the 10-, 20-, and 40-day bootstrap block lengths.
\item Performance across $B\in\{100,250,500,1000\}$ is nonmonotone for the structured designs; additional capacity alone does not guarantee improved stress-period interval score.
\end{itemize}

\section{Data versions}

ETF11 and VIX data were downloaded on May 11, 2026, and Stock50 data on June 12, 2026. The appendix analyses use the same data versions, market chronology, and memory budget as the main analysis. Tuning choices are based on the designated validation period rather than evaluation outcomes.

\end{document}